\documentclass{PoS}

\usepackage{subcaption}
\usepackage{siunitx}

\title{Application of Deep Neural Networks to Event Type Classification in IceCube}

\ShortTitle{Application of Deep Neural Networks to Event Type Classification in IceCube}

\author{
The IceCube Collaboration\footnote{For collaboration list, see PoS(ICRC2019) 1177.}\\
{\itshape \href{http://icecube.wisc.edu/collaboration/authors/icrc19_icecube}{http://icecube.wisc.edu/collaboration/authors/icrc19\_icecube}}\\
E-mail: \email{m.kronmueller@outlook.com, theo.glauch@tum.de}
}

\abstract{
The IceCube Neutrino Observatory is able to measure the all-flavor neutrino flux in the energy range between 100 GeV and several PeV. Due to the different features of the neutrino interactions and the geometry of the detector, all high-level analyses require a selection of suitable events as a first step. However, presently, no algorithm exists that gives a generic prediction of an event's underlying interaction type. One possible solution to this is the use of deep neural networks similar to the ones commonly used for 2D image recognition. The classifier that we present here is based on the modern InceptionResNet architecture and includes multi-task learning in order to broaden the field of application and increase the overall accuracy of the result. We provide a detailed discussion of the network's architecture, examine the performance of the classifier for event type classification and explain possible applications in IceCube.\\

\vspace{4mm}
{\bfseries Corresponding authors:}
\speaker{Maximilian Kronmueller}$^{1}$, Theo Glauch$^{1}$\\
{$^{1}$ \itshape Technical University of Munich}

}

\FullConference{36th International Cosmic Ray Conference -ICRC2019-\\
		July 24th - August 1st, 2019\\
		Madison, WI, U.S.A.}

\begin{document}
\section{Introduction}\label{sec:intro}
The IceCube Neutrino Observatory is a large volume neutrino detector, which is operated deep in the ice near the geographical South Pole. With its cubic-kilometer scale it is presently the largest neutrino telescope on Earth. IceCube is able to measure the all-flavor neutrino flux in the energy range between 100 GeV and several PeV. For particle detection it exploits the production of Cherenkov light by secondary relativistic particles produced in neutrino interactions. This light is measured by Icecube's 5160 digital optical modules (DOM's). Due to the different neutrino interactions with different secondary particles, various light patterns are observed in IceCube. Hence, the identification of event types is an essential part of the IceCube data processing chain. Presently, however, no \textit{generic} classification algorithm has been developed to run online. 
Deep neural networks have been shown to be a very powerful tool to classify data given a large set of training data. In this work we introduce a deep learning event classifier based on Google's InceptionResNet architecture \cite{InceptionV1, InceptionV2V3, InceptionV4Res}.

\section{Different Event Types in IceCube} \label{ch:topologies}
\begin{figure}[ht]
	\begin{subfigure}[H]{0.23\textwidth}
		\centering
		\includegraphics[width=1.\textwidth]{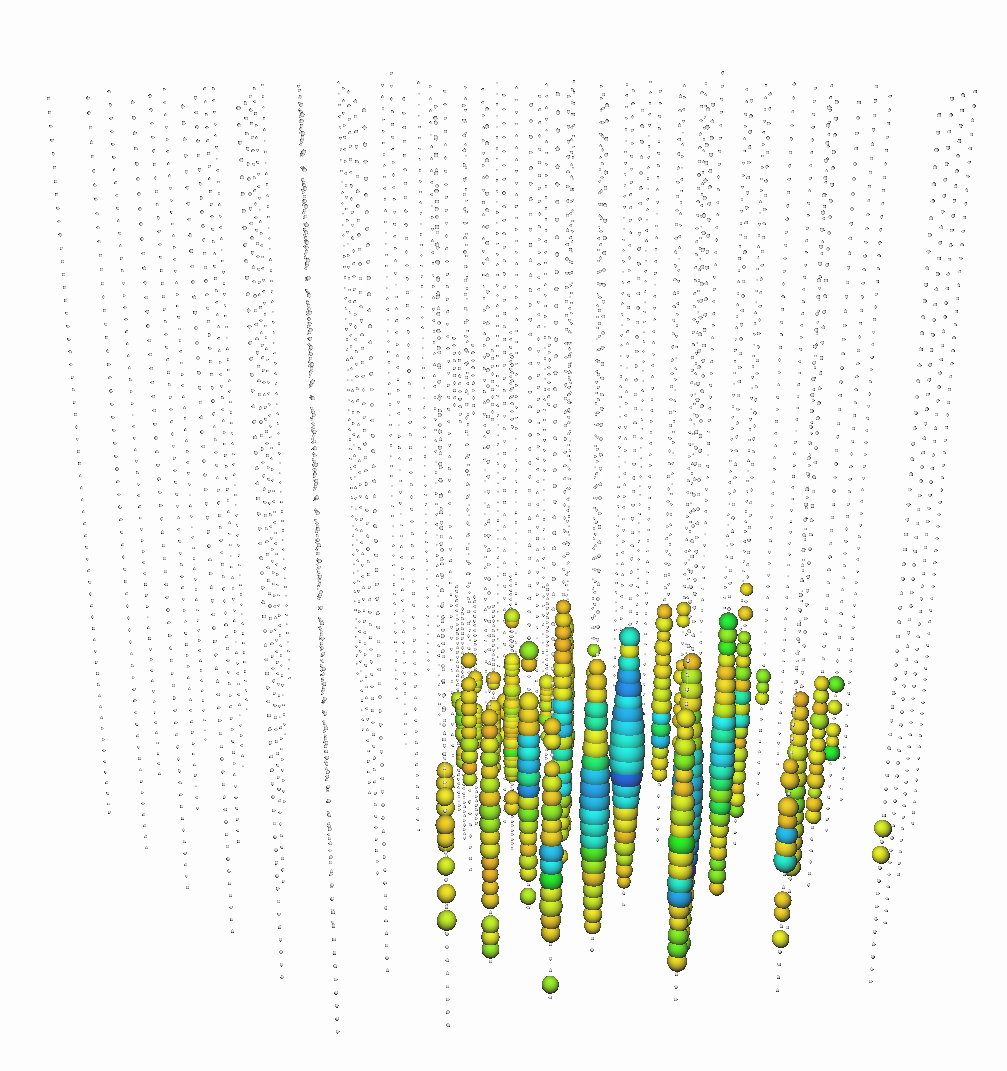}
		\caption{Cascade}
		\label{fig:CascadeEventView}
	\end{subfigure}\hfill
	\begin{subfigure}[H]{0.23\textwidth}
		\centering
		\includegraphics[width=1.\textwidth]{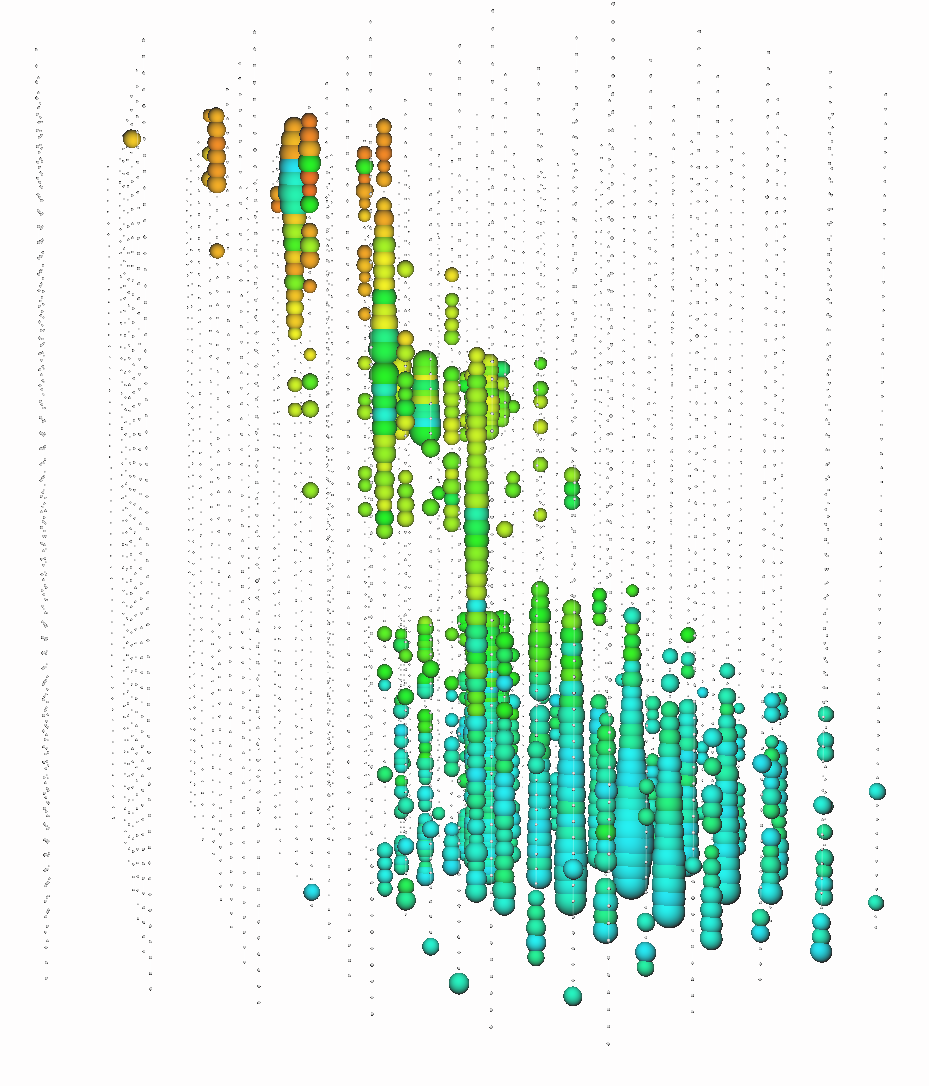}
		\caption{Track}
		\label{fig:TTrackEventView}
	\end{subfigure}
	\begin{subfigure}[H]{0.23\textwidth}
		\centering
		\includegraphics[width=1.\textwidth]{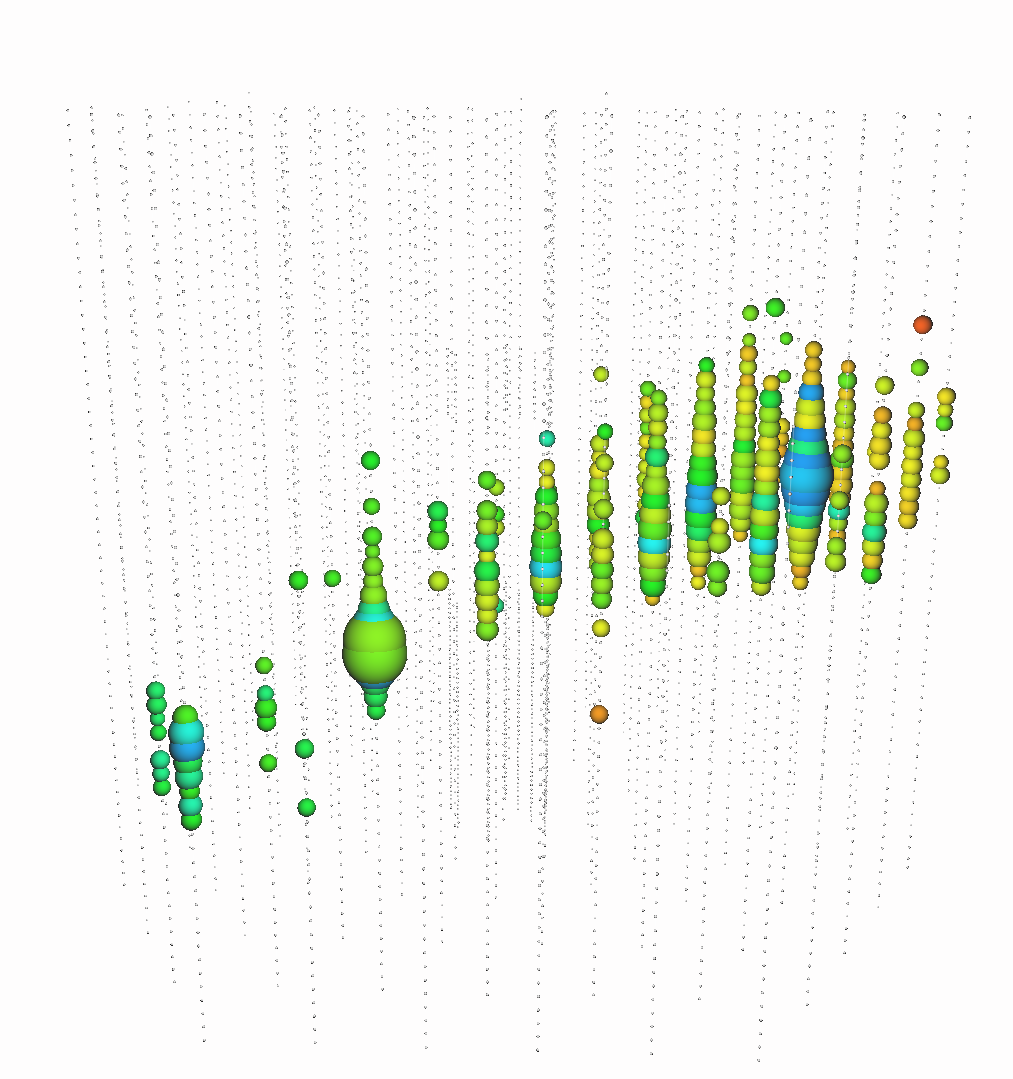}
		\caption{Starting Track}
		\label{fig:STrackEventView}
	\end{subfigure}\hfill
	\begin{subfigure}[H]{0.23\textwidth}
		\centering
		\includegraphics[width=1.\textwidth]{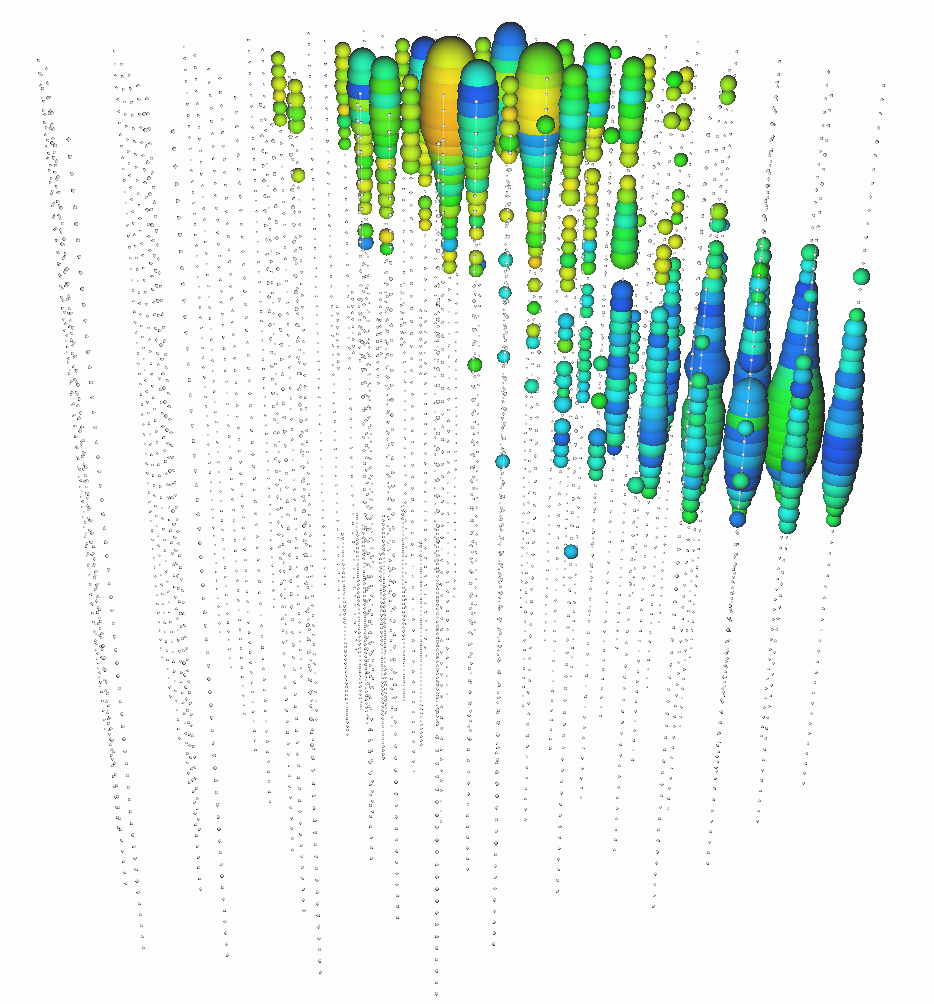}
		\caption{Double Bang}
		\label{fig:DoubleBangEventView}
	\end{subfigure}
	\caption[Event views of the main event types in IceCube]{Event views of the main event types in IceCube are shown. The detector is indicated by all DOMs drawn as small black dots. For DOMs with hits, the size of each circle correlates with the total measured charge at the DOM. The color symbolises the relative arrival times of the photons, with early hit DOMs coloured in red, late hit DOMs in blue. From left to right a cascade, a track, a starting track and a double bang are depicted.}
	\label{fig:EventViews}
\end{figure}

To distinguish different event types, we first need a clear definition - a ground truth - for each of them. In IceCube there are three main event types: cascades, tracks and double bangs. Additionally we want to distinguish between incoming and starting tracks, as sub-classes of tracks. Note that we ignore here events that don't directly deposit energy in the detector as passing tracks or cascades outside the detector volume. For an overview of the different event types see Figure \ref{fig:EventViews}.\\
Let us now describe the relevant event types in more detail:
\begin{itemize}
    \item \textit{Cascades} are mainly produced by two processes: charged current interactions of electron neutrinos or neutral current interactions of any neutrino flavor \cite{Halzen:2010yj}. The result is a hadronic and an electromagnetic cascade for charged current interactions and a single hadronic cascade in neutral current interactions. Cascades appear nearly spherical with a radius of only a few meters depending on the energy. An example is shown in Figure \ref{fig:CascadeEventView}.
    \item \textit{Tracks} are mainly produced by muons. Compared to electrons, radiative energy losses are strongly suppressed so that muons can travel up to several kilometers \cite{Halzen:2010yj} through the ice. The Cherenkov light produced when muons are passing the detector result in the track-like signature shown in Figure \ref{fig:TTrackEventView}. We further distinguish tracks into starting tracks and through-going tracks (simply called tracks in the rest of this proceeding), as this has been shown to be important for selections of highly-energetic astrophysical events \cite{1242856}. A starting track has its initial vertex inside the instrumented volume of IceCube (see Figure \ref{fig:STrackEventView}).
    \item \textit{Double bangs} are produced when a sufficiently energetic tau neutrino interacts via a charged current interaction, which produces a tau lepton that travels and decays inside the detector into an electromagnetic or a hadronic cascade ($\sim 82.5\%$ of the cases) \cite{TauThreeYears}. An extreme case of such an event is shown in Figure \ref{fig:DoubleBangEventView}. The more energetic the incoming tau is, the longer its decay length and thereby the separation of the two cascades. In ice the cascades have an average separation of $l = 50 \text{m}\cdot \left(\text{E}_{\tau}/\text{PeV}\right)$ \cite{TauThreeYears}. Assuming that IceCube allows for a spatial resolution of down to $5\text{m}$ we therefore hope to be able to identify tau neutrinos starting at energies of a few hundred TeV \cite{TauThreeYears}. Consequently only double bangs with a minimum track length of $5\text{m}$ are labeled as such for the training. Note that there are two features of the light deposition that are relevant here: The topology of two separated cascades (\textit{double bang}) and the \textit{double pulse} shape in the arrival time of the photons in one or more DOMs. See \cite{dp_proceeding} for the current results of a IceCube analyses searching for events with double-pulse features.
\end{itemize}{}
\section{The Neural Network}
\begin{figure}[ht]
	\begin{subfigure}[b]{0.4\textwidth}
		\centering
		\includegraphics[width=0.80\textwidth]{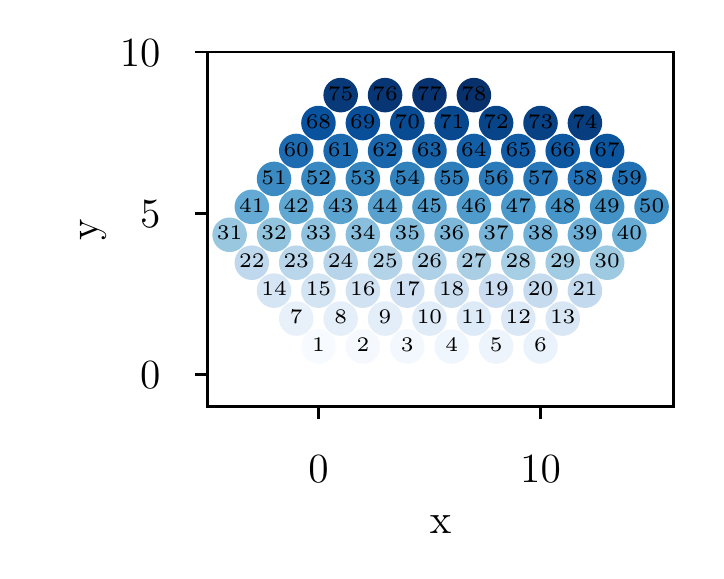}
		\includegraphics[width=0.78\textwidth]{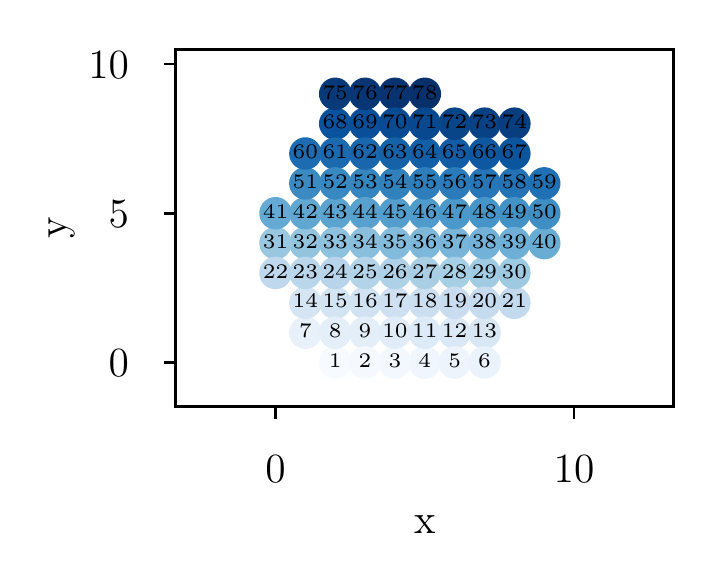}
		\caption{Grid before and after transformation}
		\label{fig:grid_icecube}
	\end{subfigure}
    \hfill
    \begin{subfigure}[b]{0.60\textwidth}
        \centering
        \includegraphics[width=1.\textwidth]{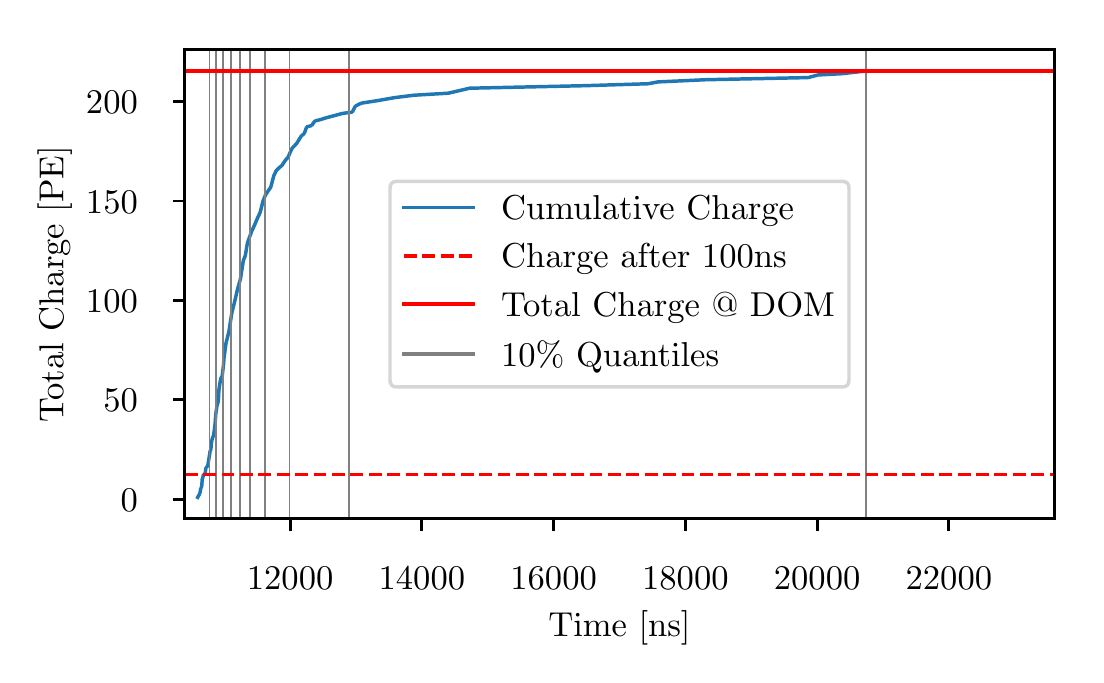}
		\caption{Visualization of the network's input features}
        \label{fig:input_features}
  \end{subfigure}
 \caption{In the left panel the IceCube grid is shown before (top) and after transformation (bottom). The original hexagonal grid is transformed on a regular grid with size $10\times 10\times 60$. Outside locations without strings are padded with zeros. In the right panel different features of a waveform measured at one specific DOM are shown. The blue line indicates the cumulative charge. A discretization of it using charge quantiles in 10\% steps is depicted by the grey lines, important areas are sampled more finely. The red lines show the charge accumulated after 100ns (dashed red line) and the total charge (solid red line).}
\label{fig:input_all}
\end{figure}

\begin{figure}[ht]
  \centering
  \includegraphics[width=.6\textwidth]{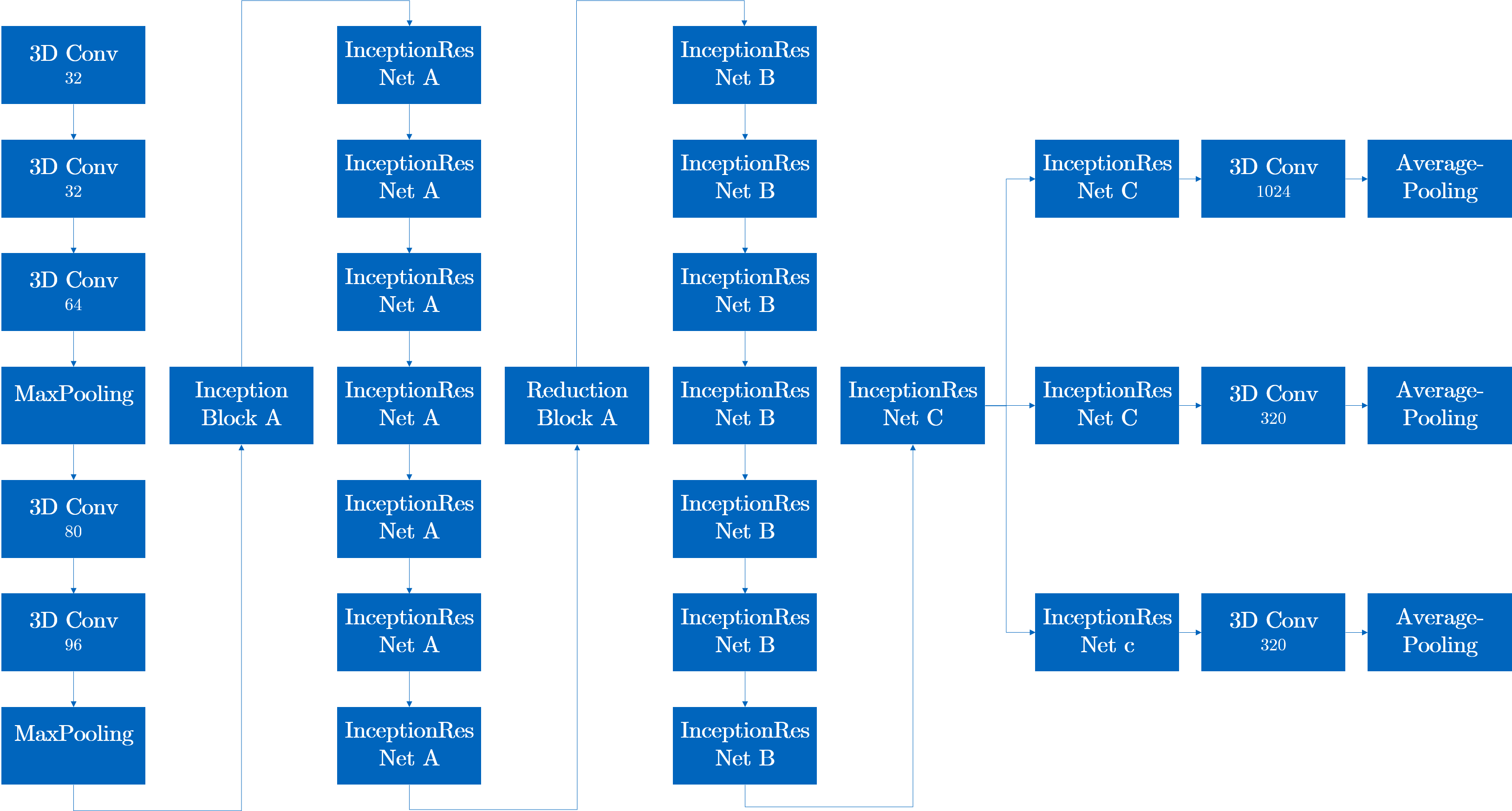}
  \caption{The general architecture of the classifier is sketched. The network consist of three main parts: the first part, a block of seven Inception-ResNet-A modules and a block of seven Inception-ResNet-B modules. They are connected by an Inception A module and later by a Reduction A block. Before the split into task-specific layers one Inception-ResNet C module is performed. For each task one further Inception-ResNet C module and one three-dimensional convolution as well as a global average pooling layer is finally added. For detailed information see \cite{InceptionV4Res}.}
  \label{fig:GlobalStrucrture}
\end{figure}

The architecture of the final neural network is based on IncResNet v2 \cite{InceptionV4Res}. As input we use a digitization of the IceCube data recorded at each DOM. At the lowest level each IceCube DOM measures a waveform - charge vs. time - which is transformed into a 'pulse series' that contains information about the total charge and time of each pulse (one or more photoelectrons). We digitize this measurement by calculating charge quantiles in different percentage steps. A sketch of the network's input features is shown in Figure \ref{fig:input_features}. By using charge quantiles instead of fixed timestamps, the areas in which the cumulative charge (blue line) changes quickly is sampled most finely and in areas where less charge is measured the digitization is coarser (see Figure \ref{fig:input_features}). For the network's input we stack the individual quantiles behind each other and append the total charge measured at each DOM. We therefore arrive at a 4-dimensional input grid: 3 spatial dimensions and one feature dimension (see also \cite{mirco_proceeding} for further discussion).
The IceCube strings are spatially distributed on a hexagonal grid. In order to convert this into a regular input, we apply a transformation that converts the original grid into a cubic grid of pixels \cite{mirco_proceeding}. The top-view of the original grid is shown in Figure \ref{fig:grid_icecube} (top). Each string is indicated and enumerated. The grid after rearranging is displayed in Figure \ref{fig:grid_icecube}. The transformation is done in a way that neighbouring relations are preserved as best as possible. Furthermore the transformation removes vacancies caused by locations with no strings in the original grid. The resulting input is a matrix of the size $10 \times 10 \times 60$ \cite{mirco_proceeding}.\\
The neural network's architecture is inspired by Google's IncResNet v2 \cite{InceptionV4Res} that combines two types of layers shown to be very powerful - Inception layers and residual units. While inception layers allow the network to learn different filters at the same time, the residual connections between the following layers stabilize the gradient in the backpropagation \cite{InceptionV4Res}. A sketch of the network's final architecture is shown in Figure \ref{fig:GlobalStrucrture}. The main difference from the original version is the transition from two-dimensional to three-dimensional convolutions. For better generalisation we perform three tasks simultaneously, hence the main branch of the network is used three times independently, once for each task-specific part. Each task adds an additional Inception-ResNet-C block, a three dimensional convolution and a global average pooling layer which is connected to the output. Between the last layer and the output we use a softmax activation function which gives a \textit{certainty score} for each class ({\it c.f.} \S\ref{sec:prob}). In this proceeding we focus on the network's main task - the event type classification, as outlined in \S\ref{ch:topologies}. The two additional output nodes perform a type-independent identification of starting events and an identification of coincident events, i.e. events with an accompanied atmospheric muon.

\section{Event Type Classification}
The dataset used for training, validation and testing of the classifier is based on Monte Carlo simulations with primary neutrino energies between 5\,TeV and 10\,PeV. 
In order to allow for stable training the training dataset is chosen in a way that minimizes the imbalance between the different event types, i.e. the same number of events from each type was pre-selected with deposited energy distribution modified such that there is an equal number of events at equal energy interval for every class. After training and validation the performance of the classifier can be accessed by looking at the test dataset which is completely independent of the training processes. An easy approach to visualize the networks performance is the use of confusion matrices. 
\subsection{Performance} \label{sch:e-to-e}
\begin{figure}[ht]
	\begin{subfigure}[h]{0.5\textwidth}
		\centering
		\includegraphics[width=1.\textwidth]{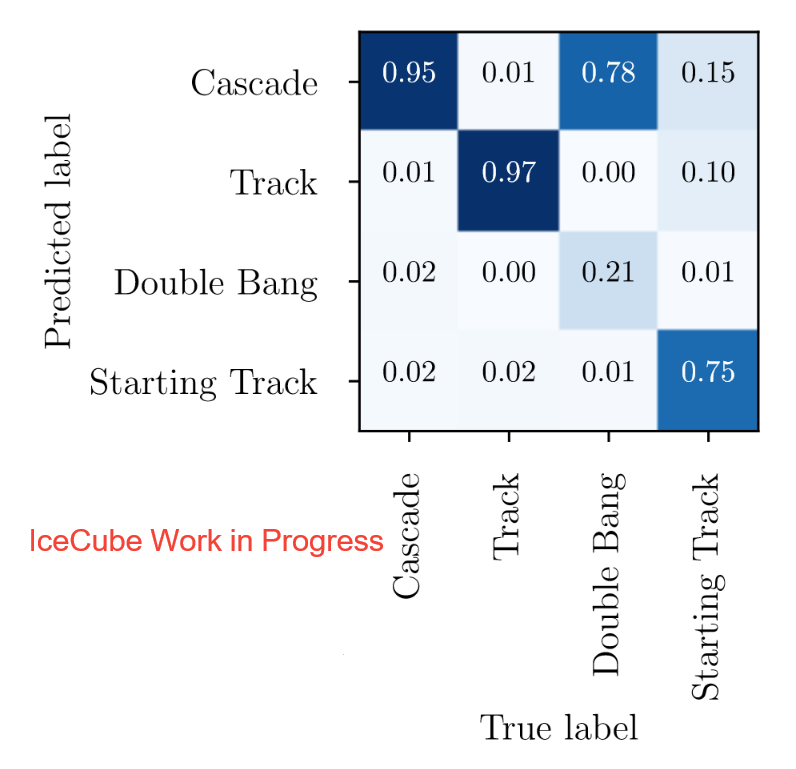}
		\caption{ground truth normalized}
		\label{fig:CM_EE_ET_all_true}
	\end{subfigure}\hfill
	\begin{subfigure}[h]{0.5\textwidth}
		\centering
		\includegraphics[width=1.\textwidth]{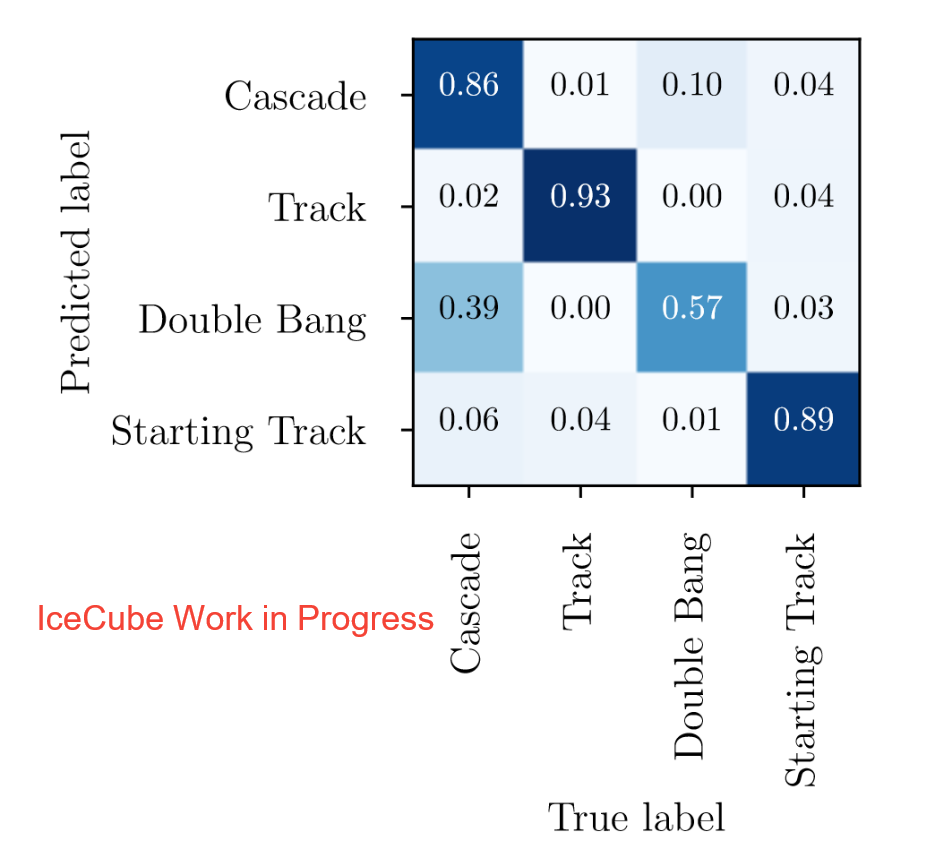}
		\caption{prediction normalized}
		\label{fig:CM_EE_ET_all_pred}
	\end{subfigure}
	\caption[Confusion matrices event type classification]{The confusion matrices of the event type classification once normalized on the ground truth of each class (left) and once normalized on the predictions of each class (right) are presented.}
	\label{fig:CM_EE_ET_all}
\end{figure}

Figure \ref{fig:CM_EE_ET_all_true} and Figure \ref{fig:CM_EE_ET_all_pred} show the confusion matrices of the event type classification normalized on the ground truth and normalized on the predictions of each class, respectively.
The network achieves best results for the track class, 97\% of them are found with a precision of 93\%. Nearly all the confusion is with starting tracks. Cascades are similarly well classified with an accuracy of 95\% and an associated precision of 86\%. While their confusion with tracks is negligible, this is not the case for double bangs and starting tracks. Consequently only 21\% of the double bang events are classified as such and a prediction of a double bang is correct in 57\% of the cases. Double bangs are often confused with cascades (78\% of the cases). 75\% of the starting tracks are identified correctly with a precision of 89\%. Starting tracks are confused with two classes: tracks and cascades.

\begin{figure}[ht]
	\begin{subfigure}[h]{0.45\textwidth}
		\centering
		\includegraphics[width=0.9\textwidth]{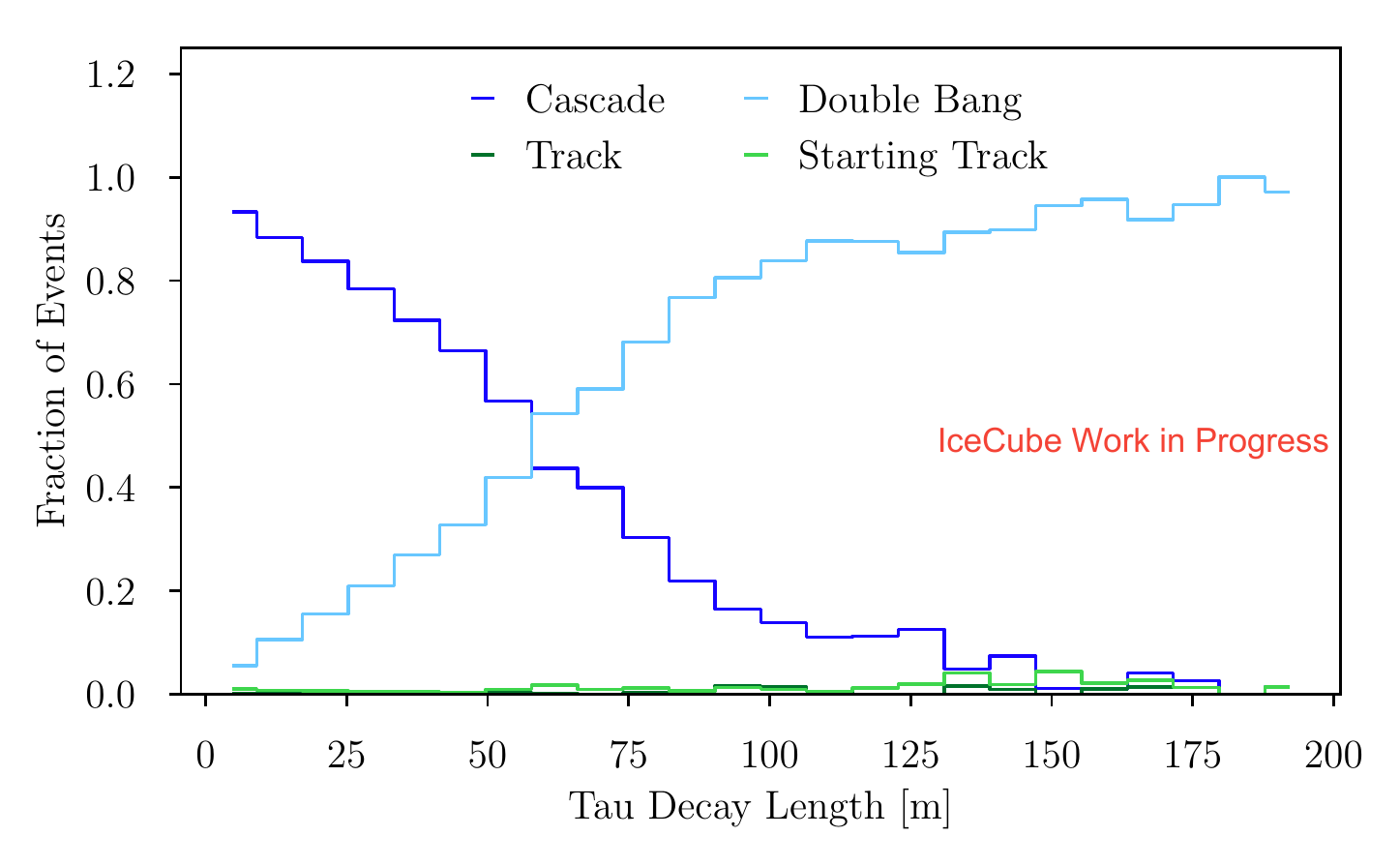}
        \caption[Predictions against tau decay length of true double bangs]{For true double bangs the prediction of each class as the fraction of all events as a function of the tau decay length is shown.}
        \label{fig:true_DB_taudecaylength}
	\end{subfigure}\hfill
	\begin{subfigure}[h]{0.45\textwidth}
	    \centering
        \includegraphics[width=0.9\textwidth]{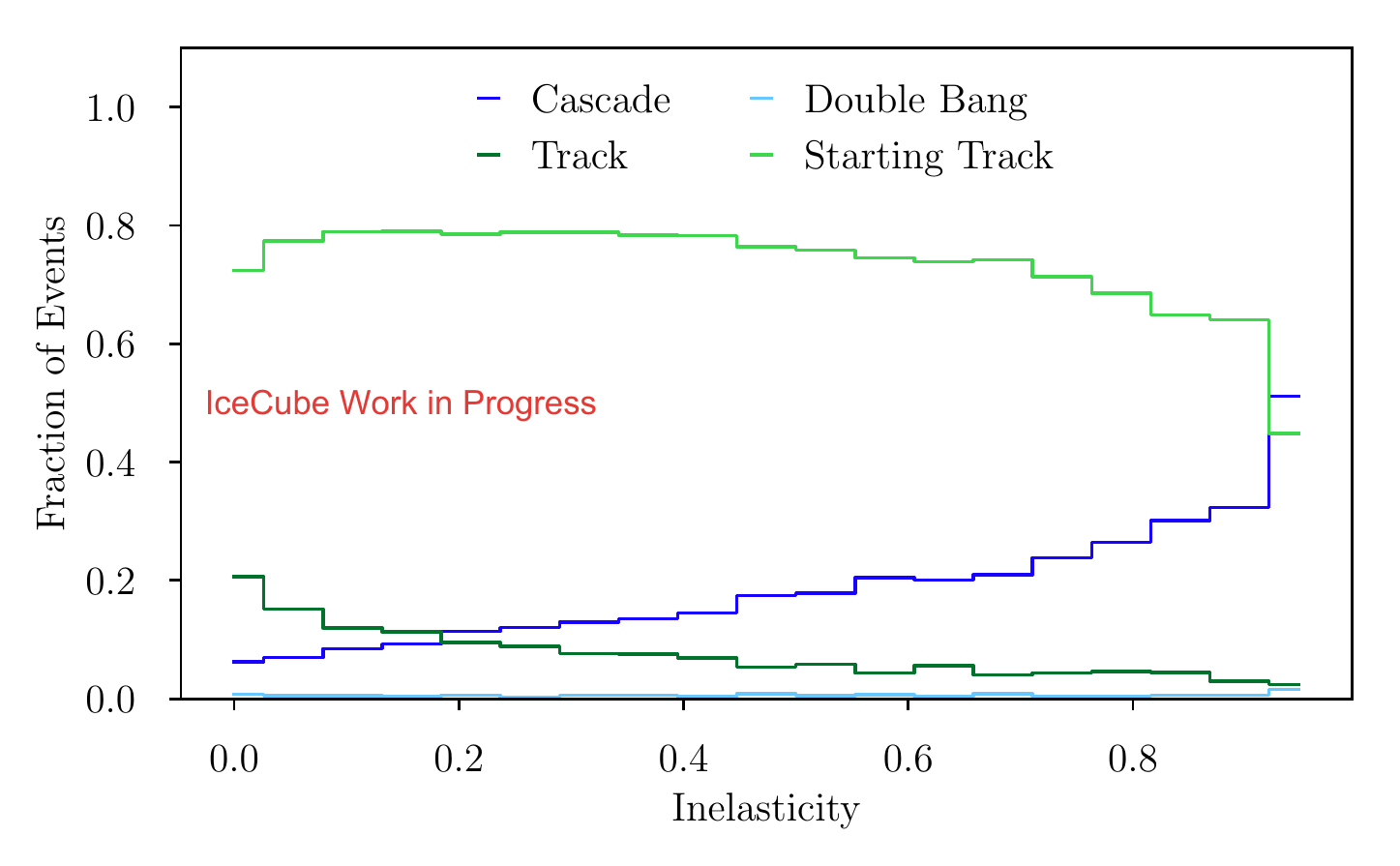}
        \caption[Predictions against inelasticity of true starting tracks]{The fraction of predictions for each class for true starting tracks as a function of the inelasticity is presented.}
        \label{fig:true_sT_inelasticity}
	\end{subfigure}
	\caption{Two exemplary tests to check the behavior of the neural network against different physical motivated principles are presented.}
	\label{fig:plausibility_tests}
\end{figure}

We next investigate the different confusions in more detail and explain their occurrence in terms of the physical expectation. With double bangs we expect to see that they are more easily predicted for longer tau decay lengths, as the two cascades are better separated from each other. In contrast, for very small tau decay lengths, we expect a confusion with cascades. In Figure \ref{fig:true_DB_taudecaylength} we show the predictions of the classifier for all \textit{true} double bangs as a function of the tau decay length. As expected the confusion with cascades dominates for short track lengths. Starting from tau decay lengths of about $60\text{m}$, however, the fraction of events predicted as cascades and the correct predictions are in balance. The confusion decreases for larger tau decay lengths until it becomes negligible for tau length above 150\,m. Double bangs show no significant confusion with either tracks or starting tracks over the whole range.

Another relevant confusion is between starting tracks and tracks and cascades. Here we expect to see a correlation between the number of correctly classified events and the inelasticity. The inelasticity $K$ of an event is defined as the ratio between the energy that goes into the hadronic cascade and the total energy of the primary neutrino.
We expect a growing confusion of starting tracks with cascades for higher inelasticities and a confusion with tracks for low inelasticities. Analogous to Figure \ref{fig:true_DB_taudecaylength} we show all events with the ground truth starting track and their corresponding predictions as a fraction of all predictions against the inelasticity in Figure \ref{fig:true_sT_inelasticity}. Again as expected, the confusion with tracks has its maximum at $K \sim 0$ and decreases with $K \rightarrow 1$. The confusion with cascades behaves inversely. It reaches its maximum of over 50\% of starting track events being wrongly classified as cascades in the last bin where nearly all the energy goes into the hadronic cascade ($K=1$). Both tendencies, a higher confusion of starting tracks with cascades for high inelasticities and a confusion with tracks for very low inelasticities, are thus confirmed.

Finally, we have compared the networks predictions to one high-level IceCube event selection, which is used for the measurement of the diffuse astrophysical neutrino flux. The sample consists of clear muon tracks with a purity of above 99\% \cite{Aartsen:2016xlq}. When comparing the network's prediction with this selection we find that the network agrees with the sample's classification in more than 99.9\% of the cases while being able to also recognize ~10\% of the (very small) cascade contamination.
\begin{figure}
  \centering

\end{figure}

\subsection{Interpretation of the Network's Output}
\label{sec:prob}
\begin{figure}[ht]
  \centering
  \includegraphics[width=0.5\textwidth]{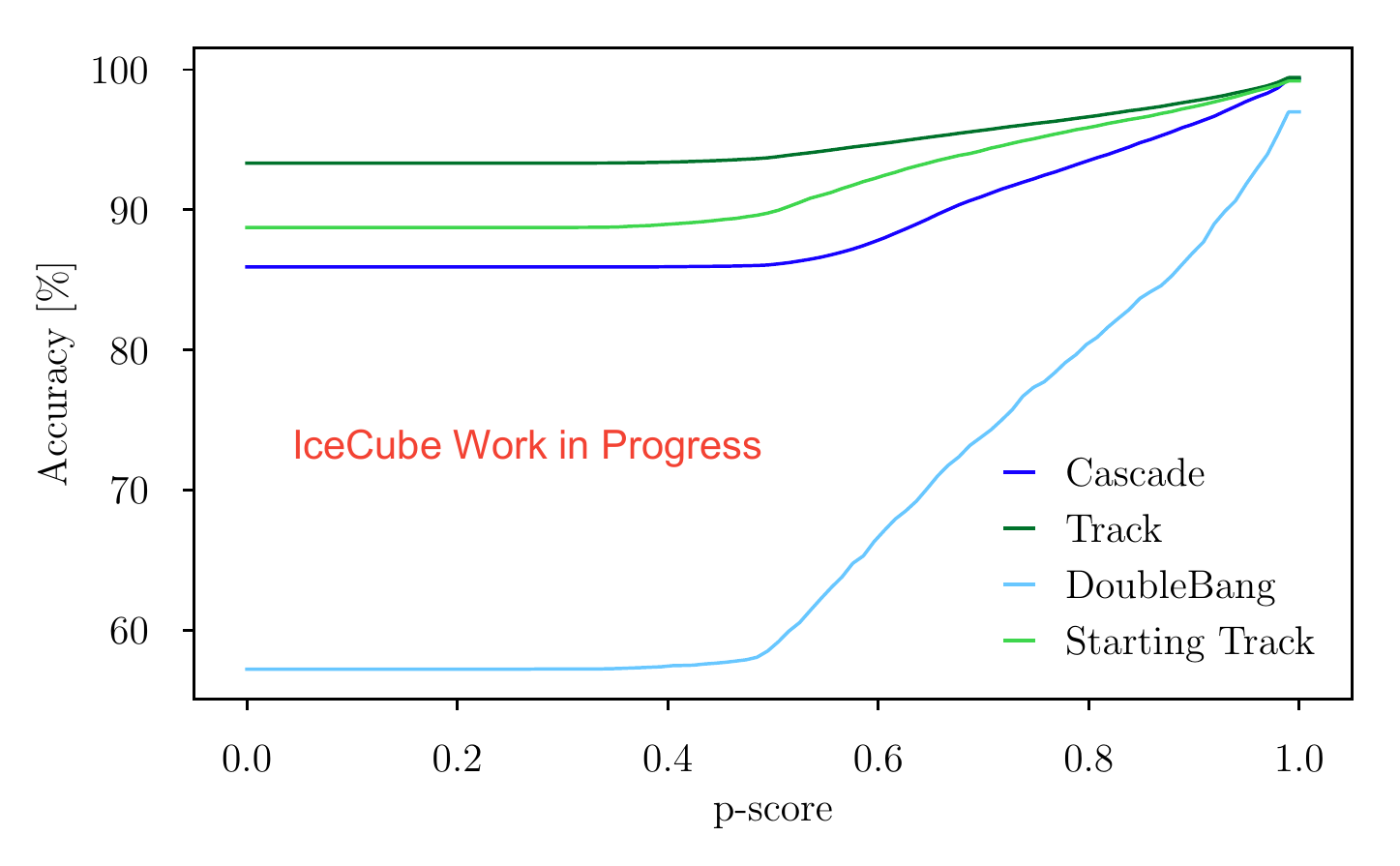}
  \caption[Accuracy against p-score cut]{The accuracy calculated for all events that are predicted with a \textit{prediction score} (p-score) higher than a certain p-score threshold is shown.}
  \label{fig:pvalue_interpretation}
\end{figure}

To interpret the prediction certainty of the neural network we use a quantity called the \textit{prediction score} (p-score). The p-score is defined as the maximum softmax output of each prediction. To use our p-score as a measure of the network's prediction certainty we require as a minimal condition that the accuracy should increase with a larger p-score threshold. Figure \ref{fig:pvalue_interpretation} shows the accuracy for each event type over the p-score threshold. It can clearly be seen that the above requirement is satisfied. Therefore we can use the p-score as a measure of certainty of the neural network's prediction. Moreover, we can see that predictions for double bang events are far less certain than for the other classes, because for the same p-scores lower accuracies are observed. Following this logic tracks are predicted to be the most certain, followed by starting tracks. All event types converge to an accuracy of 100\% for a p-score threshold of 1. The p-score can be used to select events at a desired level of purity at the price of decreased acceptance.\\

\section{Conclusion and Outlook}
In this proceeding we have shown that deep neural networks are able to classify different event types in IceCube. Confusions between the classes can be understood by physical intuition and the prediction score provides a cut variable for the purity of a selection. The long term goal is to apply such a classifier to low-level data and therefore simplify and possibly improve event selections \& filtering in IceCube. Note that due to the high speed - one prediction needs only a few ms of computing time - a classifier is also an interesting tool for real time applications. We mention here a few of the challenges and important steps for the further development:
\begin{itemize}
    \item Currently only events which deposit energy in the detector's fiducial volume are included. There is, however, a relevant background of events that are only seen by light from energy depositions outside the detector volume. A future version of the classifier is going to include these as additional class.
    \item The current training dataset only contains events above 5 TeV primary energy. IceCube is however able to measure events with (deposited) energies as low as ~100 GeV. These events will also be added to the training.
    \item Validation of the network on experimental data and CORSIKA simulations in order to understand possible disagreements between data and Monte Carlo.
\end{itemize}{}
To summarize, we are confident that Deep Learning classifications have a large potential that is still to be exploited. Besides the points mentioned above the network can also be improved to better treat irregularities in the detector geometry or include the timing through the direct usage of recurrent neural networks.


\end{document}